\documentclass[twoside,prd,superscriptaddress,showpacs,showkeys,twocolumn,nofootinbib]{revtex4}

\usepackage{amsmath}
\usepackage{amssymb}
\usepackage{slashed}
\usepackage{graphicx}
\usepackage{hyperref}
\usepackage{accents}

\newcommand{\bra}[1]{\langle#1\vert}
\newcommand{\ket}[1]{\vert#1\rangle}
\newcommand{\diag}{\mathop{\mathrm{diag}}\nolimits}
\newcommand{\I}{\ensuremath{\mathrm{i}}}
\newcommand{\e}{\ensuremath{\mathrm{e}}}
\newcommand{\eL}{\mathcal{L}}
\renewcommand{\d}{\ensuremath{\mathrm{d}}}
\newcommand{\hc}{\ensuremath{\mathrm{h.c.}}}
\newcommand{\operator}[1]{\mathop{#1}\nolimits}
\newcommand{\op}      [1]{\mathop{#1}\nolimits}

\newcommand{\group}[1]{#1}

\newcommand{\qm}[1]{``#1''} 

\newcommand{\vecleft}[1]{\accentset{\leftarrow}{#1}}
\newcommand{\vecright}[1]{\accentset{\rightarrow}{#1}}

\begin{document}

\title{Fermion flavor mixing in models with dynamical mass generation}

\author{Petr Bene\v s}
\email{benes@ujf.cas.cz}
\affiliation{Department of Theoretical Physics, Nuclear Physics Institute, \v Re\v z (Prague), Czech Republic}

\begin{abstract}
We present a model-independent method of dealing with fermion flavor mixing in the case when instead of constant, momentum-independent mass matrices one has rather momentum-dependent self-energies. This situation is typical for strongly coupled models of dynamical fermion mass generation. We demonstrate our approach on the example of quark mixing. We show that quark self-energies with a generic momentum dependence lead to an effective Cabibbo--Kobayashi--Maskawa (CKM) matrix, which turns out to be in general non-unitary, in accordance with previous claims of other authors, and to non-trivial flavor changing electromagnetic and neutral currents. We also discuss some conceptual consequences of the momentum-dependent self-energies and show that in such a case the interaction basis and the mass basis are not related by a unitary transformation. In fact, we argue that the latter is merely an effective concept, in a specified sense.
While focusing mainly on the fermionic self-energies, we also study the effects of momentum-dependent radiative corrections to the gauge bosons and to the proper vertices.
Our approach is based on an application of the Lehmann--Symanzik--Zimmermann (LSZ) reduction formula and for the special case of constant self-energies it gives the same results as the standard approach based on the diagonalization of mass matrices.
\end{abstract}



\maketitle

\section{Introduction}
\label{sec:introduction}

Flavor physics deals with two main issues: First, with the mass spectrum of leptons and quarks and second, with the coupling of leptons and quarks with different masses to one another. The latter is what we call the flavor mixing. Both issues are difficult and so far no viable explanation of their origin is known. At best, we can parameterize them by means of effective Lagrangians, valid below some energy scale. The most prominent example of these Lagrangians is that of the Standard Model of electroweak interactions (SM) \cite{Glashow:1961tr,Weinberg:1967tq,Salam:1968rm}, chiefly due to its remarkable phenomenological success (we neglect here the issue of neutrino masses and related mixing in the lepton sector).


The SM is an $\group{SU}(2)_L \times \group{U}(1)_Y$ gauge invariant theory equipped with the electroweak symmetry breaking (EWSB) sector consisting of the scalar Higgs doublet, coupled via Yukawa interactions to chiral fermions. The electroweak symmetry breakdown is driven by non-trivial vacuum expectation value (VEV) of the Higgs field, which develops due to the appropriately chosen Higgs potential. This VEV together with the Yukawa interactions give rise to fermionic mass matrices in the Lagrangian (this is what we will call the \emph{Higgs mechanism} throughout the paper). Diagonalizing these mass matrices by the bi-unitary transformation and appropriately rotating the fermionic fields from the interaction basis to the mass basis, one arrives readily at the fermion flavor mixing in the charged current sector, expressed by the unitary Cabibbo--Kobayashi--Maskawa (CKM) \cite{Cabibbo:1963yz,Kobayashi:1973fv} matrix.

Despite its phenomenological success, however, the SM suffers from various theoretical or conceptual problems. First of all, as mentioned above, it merely parameterizes the fermions masses and mixings (by the Yukawa couplings) instead of explaining them. More serious flaws, however, are connected with the EWSB sector, which is not only unmeasured so far, but it also has some serious theoretical drawbacks (especially the hierarchy problem), leading to the general opinion that the SM is only as an effective theory.

Thus, there are naturally many models on the market which go beyond the SM and try to cure (some of) its problems, connected with the mechanism of the EWSB and with the flavor physics. First of all, there are models based, unlike the SM, on some \emph{non-minimal} realization of the Higgs mechanism, in the sense that they consider other scalar content than only one doublet. One can consider two scalar doublets, resulting in the Two Higgs Doublet Model (2HDM) \cite{Lee:1973iz} or (with additional ingredients) the Minimal Supersymmetric Standard Model (MSSM) \cite{Nilles:1983ge,Haber:1984rc,Chung:2003fi}, one scalar triplet plus possibly additional scalar multiplets \cite{Schechter:1980gr,Cheng:1980qt,Gelmini:1980re}, \emph{et c{\ae}tera}.

Despite many differences, all these models have in common that they treat the flavor issues exactly in the same manner as the SM. In particular, they rely on the existence of mass matrices, generated by the Higgs mechanism, and consequently yield the same flavor mixing pattern (i.e., the unitary mixing matrix in charged current sector and no flavor changing electromagnetic and neutral currents) as the SM, using the procedure sketched above.


However, this simple picture is not the only thinkable one: The EWSB dynamics can manifest itself in the fermion sector not only by generating fermionic mass matrices. There is a plausible possibility that the EWSB dynamics generates symmetry-breaking fermionic propagators, with one-particle irreducible (1PI) parts (the proper \emph{self-energies}, sometimes also referred to as the \qm{dynamical masses}) given by flavor matrices with generic \emph{momentum dependence}. (Note that this situation in fact contains the special case of mass matrices, stemming, e.g., from the Higgs mechanism.)

This happens particularly in models of dynamical EWSB, in which the electroweak symmetry is broken down non-perturbatively by some postulated strong dynamics. This is typically accompanied by generation of the symmetry-breaking self-energies with non-trivial momentum dependence. This phenomenon is often called the \emph{dynamical} (fermion) \emph{mass generation}, since it allows to reveal the fermion masses from the dynamically generated self-energies simply by looking for the poles of the corresponding full propagators. (Note, however, that in most such models the very calculation of the momentum-dependent self-energies is a difficult task. Therefore, in reality, the self-energies are often on the basis of various arguments taken as constant, momentum-independent and accordingly regarded as the usual mass matrices.)

It must be stressed that, apart from the fermionic two-point functions, the radiatively induced momentum dependencies of other Green's functions are in principle of equal importance. In weakly coupled theories these radiative corrections can be easily (at least in principle) calculated by methods of the usual perturbation theory (once the scalars develop their symmetry-breaking VEVs). However, in strongly coupled theories this is no longer possible, since the perturbation theory does not work.

The situation is also complicated by genuine non-perturbative effect, characteristic for strongly coupled theories -- possible formation of bound states. Their omission in the internal lines of Feynman diagrams can lead to wrong results \cite{Miransky:1992da}. In order to incorporate the bound states into the calculations of Feynman diagrams, one has to know not only their spectrum (which itself is difficult to find), but also their effective couplings to the other (both elementary and composite) excitations of the theory. This can be achieved by solving Bethe--Salpeter (BS) equations \cite{Gusynin:1988cf,Appelquist:1991kn}.

\begin{figure}[t]
\begin{center}
\includegraphics[width=0.35\textwidth]{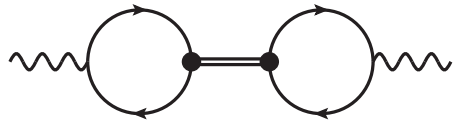}
\caption[]{A composite Nambu--Goldstone (NG) boson (double line) contribution to the gauge boson polarization tensor $\Pi_{\mu\nu}(q)$. As a fermionic composite (in this example; in principle there could be scalars as well), the NG boson couples to the fermions via momentum-dependent effective vertices (the black dots), which in turn provides an effective bilinear coupling of the NG boson to the gauge boson. The NG boson propagator thus provides the pole of the type $1/q^2$ in the polarization tensor, necessary for giving mass to the gauge boson \cite{Jackiw:1973tr}.}
\label{fig:Pi}
\end{center}
\end{figure}

In the presence of SSB, however, there is a special subclass of the bound states, whose treatment is somewhat easier: the Nambu--Goldstone (NG) bosons. First, their spectrum is known by the existence Goldstone theorem. Second, some of their (momentum-dependent) effective couplings can be relatively easily determined using the Ward--Takahashi identities \cite{Jackiw:1973tr}. These effective vertices can be in this way calculated actually only for small momenta carried by the NG boson, but this is in fact sufficient for calculating the masses of the corresponding gauge bosons, to whose polarization tensors they contribute (Fig.~\ref{fig:Pi}).

%
%

The most popular class of the models of dynamical EWSB are undoubtedly the Extended Technicolor (ETC) \cite{Dimopoulos:1979es,Eichten:1979ah} or Walking Technicolor (WTC) \cite{Holdom:1981rm,Yamawaki:1985zg} models. Roughly speaking, the EWSB is achieved in these models by new gauge (\qm{technicolor}) interactions with suitable dynamical properties (QCD-like or \qm{walking}, respectively). The corrections to the SM fermions' propagators are induced by couplings to some new postulated fermions (\qm{technifermions}), mediated by the technicolor interactions.

Not only a gauge dynamics can be responsible for the EWSB, however. Recently it was shown \cite{Benes:2006ny,Benes:2008ir} that the electroweak symmetry can be broken non-perturbatively also by sufficiently strong Yukawa interactions. The key feature of these models is that, unlike in the Higgs mechanism, the electroweak symmetry is not broken by VEVs (i.e., by one-point functions) of the scalar fields, but rather by appropriate scalar \qm{anomalous} propagators (i.e., the symmetry-breaking two-point functions). These anomalous propagators are computed self-consistently by means of the Schwinger-Dyson equations. Since the relevant dynamics is here that of Yukawa, this self-consistent treatment leads also naturally to generation of the symmetry-breaking parts, the self-energies, of the fermions. In fact, the present paper was primarily inspired by this class of models. However, there are also other papers in which strong Yukawa dynamics is employed in order to trigger the EWSB, e.g., \cite{Wetterich:2006ii,Luty:2000fj}.


Although we will deal in this paper almost exclusively with fermionic self-energies, let us, as a final example illustrating the importance of the notion of self-energies, mention the quasi-degenerate binary systems of neutral mesons. It was argued \cite{Machet:2004rv} that the standard treatment of such systems using the familiar formalism of mass matrices is inappropriate, especially when one deals with discrete symmetries like $C$, $P$, $T$ and their combinations. Instead, it was proposed to consider the appropriate matrix ($2 \times 2$) self-energies.


Noting that the concept of self-energies is crucial in a variety of models of dynamical EWSB, there is a natural question how to extract from these models the information about their flavor structure. The standard method of diagonalizing the mass matrices is clearly no longer applicable, since the self-energies cannot constitute mass matrices in any local Lagrangian due to their momentum dependence. While it is still easy to obtain the spectrum simply by looking for the poles of the full propagators, it is not immediately clear how to deal with the flavor mixing.


This question was addressed in Refs.~\cite{Appelquist:2003hn,Christensen:2005hm,Machet:2004rv,Duret:2006wk}, although in a slightly different context. The authors of Refs.~\cite{Machet:2004rv,Duret:2006wk} considered one-loop perturbative corrections to the quark propagators within the SM. However, they did not calculate these corrections explicitly, but rather assumed the resulting (finite) self-energies to be general functions of momentum. This allows to generalize their results also to non-perturbatively generated self-energies. They realized that in each fixed momentum the self-energy can constitute an \emph{approximate} mass matrix. If one chooses the momentum to be one of the pole values of the full propagator, then one of the eigenvalues of the resulting mass matrix is the physical pole mass and consequently the corresponding effective Lagrangian describes one physical state plus a number of spurious states. In this way one can construct a whole series of effective Lagrangians, which altogether describe, among a number of spurious states, all the physical states in the spectrum.

In this paper we address the same question, but following a different approach. We avoid completely the use of the mass matrices, even in an effective or approximate sense. Instead, we directly compute the observable quantities -- the amplitudes of processes involving the fermions with definite masses (i.e., flavors), obtained as the poles of the full propagators. Such a calculation can be easily performed by means of the Lehmann--Symanzik--Zimmermann (LSZ) reduction formula. In order to make a connection with the usual language of flavor physics, especially with the notions like \qm{mixing matrix} or \qm{mass basis}, we construct an effective Lagrangian which reproduces some of the amplitudes calculated by the LSZ reduction formula. Subsequently, it is possible to identify in the effective Lagrangian the quantities, which can be naturally interpreted as the effective flavor mixing matrices.

For the purposes of the present paper we restrict ourselves only to the case of quarks; application of our method to mixing of leptons, neutral mesons or other systems, should be straightforward.

Our approach reproduces and confirms the results of Refs.~\cite{Machet:2004rv,Duret:2006wk}, but from a completely different perspective. This enables us to generalize these results and interpret them in a more accurate way. In particular, we confirm that the self-energies with general momentum dependence lead to the non-unitarity of the flavor mixing matrix in the charged current sector (i.e., the effective CKM matrix) and to the flavor changing electromagnetic and neutral changing currents. However, on top of that, we find out that the ontological status of the corresponding flavor mixing matrices is merely an effective one, in the specified sense. Similarly we clarify the relation between the interaction basis and the mass basis and establish that the latter is again merely an effective notion.

Finally, in order to avoid potential confusion, it should be noted that momentum-dependent self-energies arise not only in strongly coupled theories, as suggested above. They arise in weakly coupled theories as well due to the very nature of the quantum field theory, i.e., by means of the radiative corrections. Since a perturbation theory is applicable in such a case, the standard techniques of (infinite) perturbative renormalization can be used to handle the flavor issues. This was done for the SM (in one-loop) by various authors (see, e.g., Refs.~\cite{Barroso:2000is,Diener:2001qt,Zhou:2003te,Kniehl:2006rc} and references therein). Note, however, that we consider in this paper rather non-perturbative corrections to the quarks propagators. In this case the usual perturbative renormalization machinery is inapplicable and another methods must be used, which is the aim of this paper.

The paper is organized as follows: First we introduce in Sec.~\ref{sec:formalism} some formalism concerning the momentum-dependent fermionic self-energies and diagonalizing the corresponding full propagators. In Sec.~\ref{sec:assumtions} we state the basic physical assumptions under which we work in the subsequent sections. Next, before going to the most general case of momentum-dependent self-energies, we warm-up in Sec.~\ref{sec:SM} by reminding the familiar case of the constant self-energies -- the mass matrices, known from the SM. The core of the paper is in Sec.~\ref{sec:general_selfenergies}, where we investigate the fermion flavor mixing in the case of quark self-energies with general momentum dependence. We consider in more detail the charged current sector and derive appropriately defined effective CKM matrix. As for the electromagnetic and neutral-current sectors, we do not go into a detail, since the procedure would be much the same as in the charged current sector. We merely state the results which imply a general flavor changing quark mixing.
Having investigated the effects of quark self-energies' momentum dependence, we redo in Sec.~\ref{sec:other_momentum_dep} the same analysis for other Green's function of interest, namely for the gauge boson polarization tensors and for the proper vertices of the quarks and gauge bosons.
In Sec.~\ref{sec:summary_discussion} we discuss the results obtained in previous sections. The last Sec.~\ref{sec:conclusion} is dedicated to summary and conclusion.

\section{Formalism and preliminaries}
\label{sec:formalism}

Before making any physics, we find useful to introduce in this section some notation and formul{\ae} concerning the fermionic self-energies (and corresponding full propagators) with general momentum dependence. Since we are interested in this paper in flavor mixing, we focus on the case of more fermion flavors, in which case the self-energy is a matrix in the flavor space and as such it is subject to a diagonalization.

Let there be $n$ fermions (flavors) $\psi_i$, $i=1,\ldots,n$, organized into the $n$-plet $\psi=(\psi_1,\ldots,\psi_n)^T$. Then the inverse of its full propagator
\begin{eqnarray}
\label{eq:def:S_fourier}
\bra{0}T\operator{\psi}(x)\operator{\bar\psi}(y)\ket{0}&=& \int\!\frac{\d^4p}{(2\pi)^4} \, \I S(p)\,\e^{-\I p\cdot(x-y)}
\end{eqnarray}
in the momentum representation has the most general form
\begin{eqnarray}
\label{eq:definition_S_gen}
S^{-1}(p)&=& \slashed{p}\Big(A(p^2)+\gamma_5B(p^2)\Big) - \Big(C(p^2)+\gamma_5D(p^2)\Big)
\,, \nonumber \\ &&
\end{eqnarray}
where $A$, $B$, $C$, $D$ are in principle arbitrary $p^2$-dependent $n \times n$ matrices in the flavor space. 

Within the present paper we will not consider the fermionic propagator \eqref{eq:definition_S_gen} in the full generality but rather make two approximations: First, we set $A+\gamma_5B=1$ and correspondingly neglect the wave function renormalization. This approximation does not have any substantial qualitative influence on our results. In fact, taking the wave function renormalization into account would only induce the appearance of various $Z$-factors in the formul{\ae} to come. These $Z$-factors would not alter the main message of the paper and hence we dismiss them for the sake of clarity.

Second, later on we will be using the LSZ reduction formula, which in the form, which we will use, holds only for stable states (i.e., the states which exist as asymptotic states). Hence, we demand that $S(p)=\bar S(p) \equiv \gamma_0 S^\dag(p) \gamma_0$, which ensures that the propagator $S(p)$ has poles only at real and positive $p^2$. This assumption is obviously questionable in the real world, however we make it, since still we will be able to arrive at some interesting and plausible results. The effects of finite fermion widths are subject to further research.

As a result of the two approximations, the propagator \eqref{eq:definition_S_gen} can be rewritten in the compact form
\begin{equation}
\label{eq:S}
S^{-1}(p) = \slashed{p} -  \hat\Sigma(p^2)
\,,
\end{equation}
where we introduced the notation
\begin{eqnarray}
\label{eq:definition_hatSigma}
\operator{\hat\Sigma(p^2)} &\equiv& \operator{\Sigma(p^2)} P_L + \operator{\Sigma^\dag(p^2)} P_R
\,.
\end{eqnarray}
In this expression the $\Sigma$ is an arbitrary $p^2$-dependent $n \times n$ matrix (the correspondence with Eq.~\eqref{eq:definition_S_gen} is given by $\Sigma^\dag,\Sigma = C \pm D$) and $P_{R,L}$ are the usual chiral projectors
\begin{equation}
P_{L}=\frac{1-\gamma_5}{2} \,,
\quad \quad
P_{R}=\frac{1+\gamma_5}{2} \,.
\end{equation}
It is interesting to see that Eq.~\eqref{eq:S} can be explicitly inverted as
\begin{eqnarray}
\label{eq:S_gen_inverted}
S(p) &=& \hphantom{+\,}
\Big(\slashed{p} + \Sigma^\dag(p^2)\Big) \Big(p^2-\Sigma(p^2)\,\Sigma^\dag(p^2)\Big)^{-1}P_L
\nonumber \\&& + \,
\Big(\slashed{p} + \Sigma(p^2)\Big) \Big(p^2-\Sigma^\dag(p^2)\,\Sigma(p^2)\Big)^{-1}P_R
\,.
\end{eqnarray}
We will not use this expression in the following, we only note that it is easy to deduce from it the pole equation for revealing the spectrum, which reads
\begin{equation}
\label{eq:pole_eq_nondiag}
\det \! \Big(p^2-\Sigma(p^2)\,\Sigma^\dag(p^2)\Big) = 0
\,.
\end{equation}
(Note that this equation is the same for both terms in \eqref{eq:S_gen_inverted}, since $\det(p^2-\Sigma\Sigma^\dag)=\det(p^2-\Sigma^\dag\Sigma)$.)

The next step is the diagonalization of the propagator $S(p)$. Using the \emph{bi-unitary transformation} (which is a special case of the more general \emph{singular value decomposition}) we can write the matrix $\Sigma$ in the form
\begin{equation}
\label{eq:definition_UV}
\Sigma(p^2) = U^\dag(p^2) \, M(p^2) \, V(p^2) \,,
\end{equation}
where $U$, $V$ are some unitary matrices and $M$ is a diagonal, real, non-negative matrix:
\begin{equation}
\label{def:M_diag_nonconst}
M(p^2) = \diag\Big(M_1(p^2),M_2(p^2), \ldots, M_{n}(p^2)\Big) \,.
\end{equation}
It is convenient to introduce unitary matrix
\begin{eqnarray}
X(p^2) &\equiv& V^\dag(p^2)\, P_L + U^\dag(p^2)\, P_R \,,
\end{eqnarray}
since it allows to write more compact formul{\ae}, without the necessity to use explicitly the chiral projectors $P_{L/R}$. It can be used to diagonalize $\hat\Sigma$ as
\begin{eqnarray}
\hat\Sigma(p^2) &=& \bar X^\dag(p^2) \, M(p^2) \, X^\dag(p^2) \,,
\end{eqnarray}
where $\bar X \equiv \gamma_0 X^\dag \gamma_0$. Then the propagator $S(p)$ can be diagonalized as
\begin{eqnarray}
\label{eq:S_diagonalized}
S(p) &=&  X(p^2) \frac{\slashed{p}+M(p^2)}{p^2-M^2(p^2)} \bar X(p^2) \,.
\end{eqnarray}
(This expression is correct, since the matrices in the nominator and denominator commute with each other, as they both are flavor-diagonal.)

Let us now reveal the spectrum by looking for the poles of diagonalized propagator $S(p)$. Clearly, the poles can only be found in the denominator of \eqref{eq:S_diagonalized}, i.e., by solving the pole equation
\begin{equation}
\label{eq:pole_eq_general}
\det \! \Big( p^2 - M^2(p^2) \Big) = 0 \,,
\end{equation}
which is just a diagonalized form of Eq.~\eqref{eq:pole_eq_nondiag}. This equation decouples, due to the diagonality of $M(p^2)$, into $n$ partial pole equations
\begin{equation}
\label{eq:pole_eq_partial}
p^2 - M_i^2(p^2) = 0 \quad\quad (i=1,\dots,n) \,.
\end{equation}
We will assume in the following that each partial pole equation \eqref{eq:pole_eq_partial} has exactly one solution $p^2=m_i^2$, which is necessarily non-negative due to reality of $M(p^2)$. This assumption that the full pole equation \eqref{eq:pole_eq_general} has as many poles as there are flavors (actually even stronger assumption, that each partial pole equation \eqref{eq:pole_eq_partial} has exactly one solution) is our only constraint on the otherwise arbitrary self-energy $\Sigma(p^2)$ and clearly is not crucial. In principle, nothing protects us from allowing each partial pole equation \eqref{eq:pole_eq_partial} to have $n_i$ solutions, with $n_i$ being arbitrary natural number (including 0), perhaps constraint only by some phenomenological requirement on the total number of the poles $\sum_{i=1}^n n_i$. We make this assumption only for the sake of simplicity and also in order to make a connection with the usual case of constant $\Sigma$, in which case the assumption holds.

In the following sections we will make use of the asymptotic relations for the propagator $S(p)$ with the momentum going on-shell. These relations can be easily derived under the assumption made in the previous paragraph and using the diagonal form of the propagator $S(p)$ \eqref{eq:S_diagonalized}. One gets: \footnote{There is no summation over the flavor index $i$. Any summations over the flavor indices will be always denoted explicitly throughout the paper. (Summation convention for other type of indices, e.g., the Lorentz indices, remains in use.)}
\begin{subequations}
\label{eq:S_asymptotics}
\begin{eqnarray}
S(p) & \xrightarrow[p^2 \rightarrow m_i^2]{} & \frac{\operator{\mathcal{U}_i}(p)\operator{\mathcal{\bar U}_i}(p)}{p^2-m_i^2} + \mbox{regular terms} \,, \quad
\\
S(-p) & \xrightarrow[p^2 \rightarrow m_i^2]{} & -\frac{\operator{\mathcal{V}_i}(p)\operator{\mathcal{\bar V}_i}(p)}{p^2-m_i^2} + \mbox{regular terms} \,, \quad\quad
\end{eqnarray}
\end{subequations}
where we denoted
\begin{subequations}
\label{eq:definition_mathcal_UV}
\begin{eqnarray}
\mathcal{U}_i(p) &\equiv&  \operator{X(m_i^2)} e_i \operator{u_i(p)} \,,
\\
\mathcal{V}_i(p) &\equiv&  \operator{X(m_i^2)} e_i \operator{v_i(p)}
\end{eqnarray}
\end{subequations}
(more on the interpretation of these symbols is discussed below in the subsection \ref{subsec:interpretation_U_V}) and their Dirac conjugate defined in the usual way as $\mathcal{\bar U}=\mathcal{U}^\dag\gamma_0$, $\mathcal{\bar V}=\mathcal{V}^\dag\gamma_0$. Here $e_i$ is the $i$'th canonical basis vector of $n$-dimensional flavor vector space, i.e., with the $j$'th component given by $(e_i)_j=\delta_{ij}$. Symbols $u_i(p)$, $v_i(p)$ are the standard bispinor solutions to the momentum-space Dirac equation \footnote{Note that we suppress the polarizations indices in Eqs.~\eqref{eq:def:u_v} as well as sum over them in Eqs.~\eqref{eq:S_asymptotics}. This suppression will be carried on systematically throughout the paper, in the case of fermions as well as in the case of vector bosons.}
\begin{subequations}
\label{eq:def:u_v}
\begin{eqnarray}
(\slashed{p}-m_i)\operator{u_i(p)} &=& 0 \,,
\\
(\slashed{p}+m_i)\operator{v_i(p)} &=& 0 \,.
\end{eqnarray}
\end{subequations}

Having defined the momentum-dependent matrices $V(p^2)$, $U(p^2)$ (Eq.~\eqref{eq:definition_UV}), it will be useful in the following to define their momentum-independent counterparts $\tilde V$, $\tilde U$, such that their elements on position $i,j$ are defined as
\begin{subequations}
\label{eq:def:U_V_tilde}
\begin{eqnarray}
(\tilde V)_{ij} &=&  (V(m_i^2))_{ij} \,,
\\
(\tilde U)_{ij} &=&  (U(m_i^2))_{ij} \,,
\end{eqnarray}
\end{subequations}
i.e., explicitly
\begin{equation}
\tilde V =
\left(
  \begin{array}{cccc}
    V_{11}(m_1^2) & V_{12}(m_1^2) & \cdots & V_{1n}(m_1^2) \\
    V_{21}(m_2^2) & V_{22}(m_2^2) &        & V_{2n}(m_2^2) \\
    \vdots        &               & \ddots & \vdots        \\
    V_{n1}(m_n^2) & V_{n2}(m_n^2) & \cdots & V_{nn}(m_n^2) \\
  \end{array}
\right)
\end{equation}
and similarly for $\tilde U$. We can also define the constant matrix $\tilde X$ as
\begin{eqnarray}
\label{eq:def:X_tilde}
\tilde X &\equiv& \tilde V^\dag P_L + \tilde U^\dag P_R \,.
\end{eqnarray}
Obviously, for constant (momentum-independent) $U$, $V$ we have $\tilde V=V$, $\tilde U=U$ and consequently $\tilde X = X$. In this case the matrices $\tilde V$, $\tilde U$ and $\tilde X$ are also unitary, which need not be true in general.

\subsection{Interpretation of the $\mathcal{U}$, $\mathcal{V}$ symbols}
\label{subsec:interpretation_U_V}

Let us add a comment on how to interpret the symbols $\mathcal{U}_i$, $\mathcal{V}_i$. Assume in this subsection that the self-energy $\Sigma$ is a \emph{constant} (i.e., momentum-independent) matrix, i.e., effectively a mass matrix in the Lagrangian. Then the plane-wave solutions to the Dirac equation $(\I\slashed{\partial}-\hat\Sigma)\psi = 0$
with positive and negative energy (we assume $p_0>0$) read
\begin{subequations}
\begin{eqnarray}
\psi_{+}(x) &=& \mathcal{U}(p) \, \e^{-\I p \cdot x} \,,
\\
\psi_{-}(x) &=& \mathcal{V}(p) \, \e^{+\I p \cdot x} \,,
\end{eqnarray}
\end{subequations}
where the quantities $\mathcal{U},\mathcal{V}$ satisfy
\begin{subequations}
\begin{eqnarray}
(\slashed{p}-\hat\Sigma)\,\mathcal{U}(p) &=& 0 \,,
\\
(\slashed{p}+\hat\Sigma)\,\mathcal{V}(p) &=& 0 \,.
\end{eqnarray}
\end{subequations}
Now using $\hat\Sigma=\bar X^\dag M X^\dag$, with $M=\diag(m_1,\ldots,m_{n})$, we arrive at
\begin{subequations}
\begin{eqnarray}
\mathcal{U}(p) &=& \sum_i X \, e_i \operator{u_i}(p) \>\equiv\> \sum_i \mathcal{U}_i(p) \,,
\\
\mathcal{V}(p) &=& \sum_i X \, e_i \operator{v_i}(p) \>\equiv\> \sum_i \mathcal{V}_i(p) \,,
\end{eqnarray}
\end{subequations}
which (for momentum-independent $X$) coincides with definitions \eqref{eq:definition_mathcal_UV}. Thus, we can understand the symbol $\mathcal{U}_i(p)$ ($\mathcal{V}_i(p)$) as the polarization vector of the fermion (antifermion) of $i$'th flavor with mass $m_i$, as a generalization of the usual polarization vector $u_i(p)$ ($v_i(p)$) in the case of multicomponent fermionic field $\psi$.

\section{Setting the stage}
\label{sec:assumtions}



Concerning the physical context, we make two principal assumptions:


\subsection{Symmetric Lagrangian}

First, we assume that there is an $\group{SU}(2)_L \times \group{U}(1)_Y$ invariant Lagrangian $\eL(u^\prime,d^\prime)$, containing $n$ generations (flavors) of quark fields, organized in the usual $\group{SU}(2)_L \times \group{U}(1)_Y$ multiplets. We deliberately put the up-type and down-type quark fields into the flavor $n$-plets
\begin{equation}
\label{eq:def:primed_quarks}
u^\prime = \left( \begin{array}{c} u_1^\prime \\ u_2^\prime \\ \vdots \\ u_n^\prime \end{array} \right)
\,,
\quad\quad\quad
d^\prime = \left( \begin{array}{c} d_1^\prime \\ d_2^\prime \\ \vdots \\ d_n^\prime \end{array} \right)
\,,
\end{equation}
(we combine here the chiral quark fields as $q^\prime=q_L^\prime+q_R^\prime$, $q^\prime=u^\prime,d^\prime$). Thus, in essence, $\eL(u^\prime,d^\prime)$ is (up to the absence of leptons) nothing else than the usual SM Lagrangian \emph{without} neither the Higgs sector nor any other EWSB sector. In particular, it contains (apart from the gauge self-couplings) only the gauge interactions of charged, neutral and electromagnetic currents, which we state here for the sake of later references:
\begin{subequations}
\label{eq:L_gauge}
\begin{eqnarray}
\eL_{\mathrm{cc}}(u^\prime,d^\prime) &=& \frac{g}{\sqrt{2}}\bar u^\prime \gamma^\mu P_L d^\prime W_\mu^\dag + \hc \,,
\label{eq:L_gauge_cc}
\\
\eL_{\mathrm{nc}}(u^\prime,d^\prime) &=& \frac{g}{2 \cos \theta_W}
\sum_{q^\prime=u^\prime,d^\prime}
\bar q^\prime \gamma^\mu (v_q-a_q\gamma_5) q^\prime Z_\mu \,,
\nonumber \\ &&
\\
\eL_{\mathrm{em}}(u^\prime,d^\prime) &=& \sum_{q^\prime=u^\prime,d^\prime} e Q_q
\bar q^\prime \gamma^\mu q^\prime A_\mu \,.
\end{eqnarray}
\end{subequations}
We use the standard notation
\begin{subequations}
\begin{eqnarray}
v_q &=& T_q^3-2Q_q \sin^2\theta_W \,,
\\
a_q &=& T_q^3 \,,
\end{eqnarray}
\end{subequations}
where $T_q^3$ is the third component of the weak isospin and $Q_q$ is the electric charge of the corresponding quark fields, i.e.,
\begin{subequations}
\begin{align}
T_u^3 &= +\frac{1}{2} \,, &
Q_u   &= +\frac{2}{3} \,,
\\
T_d^3 &= -\frac{1}{2} \,, &
Q_d   &= -\frac{1}{3} \,.
\end{align}
\end{subequations}
According to the common convention in the literature, we will sometimes refer to the fields $u^\prime$, $d^\prime$ as the \emph{(weak) interaction basis}, since the interaction Lagrangian \eqref{eq:L_gauge} written in their terms is flavor-diagonal.

A remark concerning the denotation of the interaction basis is in order here. The primes at \eqref{eq:def:primed_quarks} indicate that the fields in question do not correspond at this stage to the physical (massive) quarks. Later on we will (in various contexts) introduce unprimed quark fields $u$, $d$, corresponding directly to the massive quarks and hence referred to as the \emph{mass basis}. On the other hand, since the massiveness of gauge bosons is not of our primary interest, we will not be so careful in distinguishing the massless and massive gauge bosons (and the corresponding operators). We will denote them all by the same symbol $W^\pm$, $Z$, the actual meaning of which should be always clear from the context.


The leptons could be introduced as well. It would be interesting especially upon considering the massive neutrinos, since then there are richer possibilities concerning the structure of their self-energy, due to the possibility of neutrinos being Majorana particles. However, this richness, as compared to the case of quarks, might obscure the idea we want to explain and that is why we restrict our discussion only to the quarks. Possible application to the leptons should be straightforward, although more tedious.



\subsection{Symmetry-breaking dynamics}

Second, we assume that there is some EWSB dynamics beyond $\eL(u^\prime,d^\prime)$, which breaks the $\group{SU}(2)_L \times \group{U}(1)_Y$ symmetry down to $\group{U}(1)_{em}$. This is manifested by generation of symmetry-breaking corrections to various Green's functions, in particular, to the quark and gauge boson propagators and to their vertices.

The point is that in order to keep our analysis as general as possible, we do not specify this model-dependent dynamics in detail. The best what can be done in this situation with a lack of knowledge about the dynamics is merely to parameterize somehow the effects of the dynamics on the Green's functions of interest. Thus, we will consider the radiative corrections as the most general functions of involved momenta, consistent with Lorentz symmetry.

If the dynamics in question is strong, the formation of bound states is to be expected. Again, without knowing the precise details of given theory their treatment is rather difficult. Among various models there are substantial differences not only in the dynamics of the elementary constituents of the bound states, but also in the very spectrum of the elementary constituents. For example, in technicolor theories the NG bosons are composite fields made of fermions. On the other hand, there are models (e.g., \cite{Benes:2006ny,Benes:2008ir,Wetterich:2006ii,Luty:2000fj}) with dynamical EWSB, in which the NG bosons are bound states of not only fermions, but also of some elementary scalars. All these differences clearly affect the Bethe--Salpeter-based treatment of the bound states.

For these reasons we will not consider explicitly the effective couplings of bounds states. Although they could be treated (i.e., parameterized) in a similar way as the vertices of quarks and gauge bosons, we will not do it, since we actually do not need it. That is because we will never need to consider on-shell bound states (i.e., as external legs in Feynman diagrams). Nevertheless, the bound states will be included in our analysis implicitly, through their contributions to the radiative corrections mentioned above.


The radiative corrections to the quark propagators are assumed to be of the form considered in Sec.~\ref{sec:formalism}. Thus, without going into detail, we only state here, for the purpose of later references, the resulting EWSB quark propagators $S_u(p)=\langle u^\prime \bar u^\prime \rangle$, $S_d(p)=\langle d^\prime \bar d^\prime \rangle$:
\begin{subequations}
\label{eq:S_u_d}
\begin{eqnarray}
S_u^{-1}(p) &=& \slashed{p} - \Big( \Sigma_u(p^2) \, P_L + \Sigma_u^\dag(p^2) \, P_R \Big) \,,
\\
S_d^{-1}(p) &=& \slashed{p} - \Big( \Sigma_d(p^2) \, P_L + \Sigma_d^\dag(p^2) \, P_R \Big) \,.
\end{eqnarray}
\end{subequations}
Recall that the matrices $\Sigma_u(p^2)$, $\Sigma_d(p^2)$ are \emph{arbitrary} complex $n \times n$ matrices, eventually constrained only by the requirement that the corresponding full propagators $S_u(p)$, $S_d(p)$ have exactly $n$ poles (for details cf. Sec.~\ref{sec:formalism}). These poles determine the mass spectrum of the theory. We will denote the corresponding states (i.e., the physical massive quarks, sometimes in this paper also referred to as the \emph{mass eigenstates}) as $u_i$, $d_i$, $i=1, \ldots, n$, with masses $m_{u_i}$ and $m_{d_i}$. If we order these states by size of their masses, this denotation is (for $n=3$) just another name for the usual up ($u_1$), charm ($u_2$), top ($u_3$) and down ($d_1$), strange ($d_2$), bottom ($d_3$) quarks.

From other Green's functions we will be interested in the gauge boson propagators and in the vertices connecting the quarks and the gauge bosons. However, their momentum-dependence will be considered only in Sec.~\ref{sec:other_momentum_dep}. In the following we restrict ourselves, for the sake of clarity, only to free (massive for $W^\pm$, $Z$ and massless for photon) vector boson propagators and to bare vertices (stemming from the interaction Lagrangian).



Note that our assumption about the existence of an EWSB sector beyond $\eL(u^\prime,d^\prime)$ is quite general in the sense that is contains a wide class of models of EWSB. In particular, it contains, as a special case, models based on some realization of the Higgs mechanism, with constant self-energies interpreted as mass matrices. However, the class of models which we will have primarily in mind throughout the paper are the models of dynamical mass generation with non-trivial momentum dependence of the corresponding self-energies (cf. Sec.~\ref{sec:introduction}).

\begin{center}
*
\end{center}

In the following we will refer to the objects of the two assumptions stated in the previous two subsections, i.e., to the symmetry-conserving Lagrangian $\eL(u^\prime,d^\prime)$ and to the symmetry-breaking full propagators $S_u(p)=\langle u^\prime \bar u^\prime \rangle$ and $S_d(p)=\langle d^\prime \bar d^\prime \rangle$, collectively as the \emph{exact theory} \footnote{We do not refer to it as the \emph{full theory} in order to minimize the risk of confusion, since the latter term would be more appropriate for something else: For the exact theory \emph{plus} the full dynamics, which generates the full propagators.}. This is in order to distinguish it from the \emph{effective theory}, to be introduced later.

\section{Standard Model}
\label{sec:SM}

Before investigating the general case of momentum-dependent self-energies in the next chapter, we revise in this short chapter how the fermion flavor mixing is treated in the special case of constant self-energies. In other words, we review here the SM; thus, this chapter can be omitted at first reading. Nevertheless, we present it here, in order to establish some notation and to make the paper self-contained. The primary reason is, however, that the SM provides a natural reference point when discussing in the next chapter some novel consequences stemming from self-energies' momentum dependence.

In the case of constant self-energies the general discussion of diagonalization of fermionic propagators in Sec.~\ref{sec:formalism} still applies, the only difference is that all the matrices $U$, $V$, $X$, $M$, $\Sigma$ are now momentum-independent. What is novel in this situation, however, is that the self-energy $\Sigma$ can be now regarded as a mass matrix sitting in the Lagrangian. In other words, the EWSB sector now generates directly mass terms in the Lagrangian. This is precisely how the Higgs mechanism works.

Hence, interpreting the self-energies as mass-matrices, we can join the two basic components of the exact theory, the self-energies $\Sigma_u$, $\Sigma_d$, defining the spectrum of the theory, and the Lagrangian $\eL(u^\prime,d^\prime)$, defining structure of its interactions, into the single Lagrangian
\begin{equation}
\label{eq:L_SM}
\eL_{\mathrm{SM}}(u^\prime,d^\prime) \equiv \eL(u^\prime,d^\prime)
- \bar u^\prime \hat \Sigma_u u^\prime - \bar d^\prime \hat \Sigma_d d^\prime
\end{equation}
(plus the suppressed gauge bosons mass terms), which now describes massive particle spectrum (for the definition of $\hat\Sigma_q$, $q=u,d$, see Eq.~\eqref{eq:definition_hatSigma}). It is identical to the Lagrangian of the SM (up to the Higgs boson interactions), hence the subscript. The whole discussion of the diagonalization of the propagators in Sec.~\ref{sec:formalism} still applies, but now it can be interpreted also as the mass diagonalization of the Lagrangian. Using the definition of the matrix $X$ (now momentum-independent) from the Sec.~\ref{sec:formalism}, we can define new fields
\begin{equation}
\label{eq:def:quark_unitary_redef}
u = X_u^\dag u^\prime \,,
\quad\quad\quad
d = X_d^\dag d^\prime \,,
\end{equation}
so that the Lagrangian \eqref{eq:L_SM} expressed now in their terms is mass-diagonal. I.e., the fields $u$, $d$ have now straightforward interpretation as fields that create the states with definite mass from the vacuum and we are allowed to call them the mass basis.~\footnote{Note that we use the same denotation $u$, $d$ for the particles and for their operators.}

As for the applying the redefinitions \eqref{eq:def:quark_unitary_redef} to the rest of the Lagrangian \eqref{eq:L_SM} (i.e., to the Lagrangian $\eL(u^\prime,d^\prime)$), we first note that the fermionic kinetic terms remain untouched, due to the unitarity of the matrices $X_u$, $X_d$. Second, upon redefining the fermionic fields also in the gauge interactions sector \eqref{eq:L_gauge}, we arrive at
\begin{subequations}
\label{eq:L_gauge_SM}
\begin{eqnarray}
\eL_{\mathrm{SM,cc}}(u,d) &=& \frac{g}{\sqrt{2}}\bar u \gamma^\mu P_L V_{\mathrm{CKM}} d W_\mu^\dag + \hc \,,
\label{eq:L_gauge_SM_cc}
\\
\eL_{\mathrm{SM,nc}}(u,d) &=& \frac{g}{2 \cos \theta_W} \sum_{q=u,d} \bar q \gamma^\mu (v_q-a_q\gamma_5) q Z_\mu \,,
\nonumber \\ &&
\label{eq:L_gauge_SM_nc}
\\
\eL_{\mathrm{SM,em}}(u,d) &=& \sum_{q=u,d} e Q_q \bar q \gamma^\mu q A_\mu \,.
\label{eq:L_gauge_SM_em}
\end{eqnarray}
\end{subequations}
We see that in contrast to the Lagrangian \eqref{eq:L_gauge}, the charged current interactions are no longer flavor-diagonal, but rather exhibit the flavor mixing parameterized by the celebrated Cabibbo-Kobayashi-Maskawa (CKM) matrix \cite{Cabibbo:1963yz,Kobayashi:1973fv}, which is expressed in terms of the matrices $V_u$, $V_d$ as
\begin{equation}
\label{eq:CKM_SM}
V_{\mathrm{CKM}} \equiv V_{u\vphantom{d}}^{\vphantom{\dag}} V_d^\dag \,.
\end{equation}
Note that $V_{\mathrm{CKM}}$ is unitary due to the unitarity of matrices $V_u$, $V_d$.
On the other hand, the electromagnetic and neutral-current interactions remain flavor diagonal, which is again a consequence of the unitarity of the matrices $X_u$, $X_d$.

The simplest physical process in which the effect of the CKM matrix takes place is the decay $W^+ \rightarrow u_i + \bar d_j$. Let us now calculate for the sake of later references its $S$-matrix element
\begin{subequations}
\label{eq:definition:Mfi}
\begin{eqnarray}
S_{fi}
&=& \bra{u_i,\bar d_j}S\ket{W^+}
\\
&=& \delta_{fi}+(2\pi)^4\,\delta^4(p+k-q)\,\I\mathcal{M}_{fi}
\end{eqnarray}
\end{subequations}
(we assign the external momenta as $W^+(q) \rightarrow u_i(p) + \bar d_j(k)$). In the lowest order in the gauge coupling constant $g$ we have immediately
\begin{equation}
\label{eq:Mfi_SM}
\mathcal{M}_{fi}^{\mathrm{(SM)}} =
\frac{g}{\sqrt{2}}\operator{\bar u_{u_i}(p)} \gamma^\mu P_L (V_{\mathrm{CKM}})_{ij} \operator{v_{d_j}(k)} \operator{\varepsilon_\mu(q)} \,.
\end{equation}
We have added the superscript $^{\mathrm{(SM)}}$ in order to indicate that the matrix element \eqref{eq:Mfi_SM} was calculated within the SM interaction Lagrangian \eqref{eq:L_gauge_SM} and in order to distinguish it from another independent calculation of the same quantity \eqref{eq:definition:Mfi} carried out in the subsequent section.

\section{Momentum-dependent self-energies}
\label{sec:general_selfenergies}


Let us now relax the requirement of the self-energies' momentum-independence and allow them to depend on momentum in a general way. In this situation the self-energies cannot be any longer interpreted as mass matrices and there is no obvious way how to reexpress the exact theory by a Lagrangian, which would be unique, local, renormalizable, and equivalent to the exact theory in the sense that it would contain the same amount of physical information as the original exact theory. However, we will show that if one releases the requirement of the full equivalence to the exact theory (as discussed in more detail in Sec.~\ref{sec:summary_discussion}), it is possible to construct an effective Lagrangian, which mimics some aspects of the exact theory.

The crucial observation is that although we do not have the exact theory expressed only by a single Lagrangian of the type $\eL(u,d)$, with $u$, $d$ being the mass eigenstates' operators (like in the previous section), but rather represented by the Lagrangian $\eL(u^\prime,d^\prime)$ \emph{plus} the full propagators $S_u(p)=\langle u^\prime \bar u^\prime \rangle$ and $S_d(p)=\langle d^\prime \bar d^\prime \rangle$, it is still possible to calculate the amplitudes of the processes involving the mass eigenstates $u$, $d$. This is allowed by the LSZ reduction formula \cite{Lehmann:1954rq}, which states that the amplitude of a given process involving the mass eigenstates $u$, $d$ can be calculated (up to the polarization vectors and possible sign due to the fermionic nature of involved particles) as a residue of the appropriate (momentum space) connected Green's function for the external momenta going on their mass-shell. The point is that the Green's function need \emph{not} be calculated in terms of eventual operators $u$, $d$ of the mass eigenstates, but rather in terms of the original interaction basis operators $u^\prime$, $d^\prime$, which have no direct connection to the mass eigenstates.
Note that the Green's functions are easily calculated in the exact theory: One can apply the usual perturbation theory given by the Lagrangian $\eL(u^\prime,d^\prime)$, with the additional Feynman rule that the fermionic lines (propagators) in the diagrams are not the bare ones, determined by the free part of $\eL(u^\prime,d^\prime)$, but rather those that are symmetry-breaking defined by Eqs.~\eqref{eq:S_u_d}.

The possibility of calculating processes involving the mass eigenstates, as sketched in the previous paragraph, opens the way to investigating the fermion flavor mixing in the case of momentum-dependent self-energies. We explain it in more detail in the following subsection on the example of flavor mixing in the charged current sector. Next, in the subsequent subsection, we state (without detailed derivation) the analogous results for the electromagnetic and neutral-current sectors. The detailed discussion of the obtained results is postponed to Sec.~\ref{sec:summary_discussion}.


\subsection{Charged current interactions}
\label{subsec:cc}

The idea is simple and can be roughly stated as follows: First, we calculate (using the approach described above) the $S$-matrix element for the process $W^+ \rightarrow u_i + \bar d_j$ in the lowest order in the gauge coupling constant. Second, we demand that the obtained $S$-matrix element has the same form as the one calculated within the SM (Sec.~\ref{sec:SM}) and define this way the effective CKM matrix. This effective CKM matrix is eventually interpreted to be a part of the effective Lagrangian of the SM form \eqref{eq:L_gauge_SM}.

\begin{figure}[t]
\begin{center}
\includegraphics[width=0.5\textwidth]{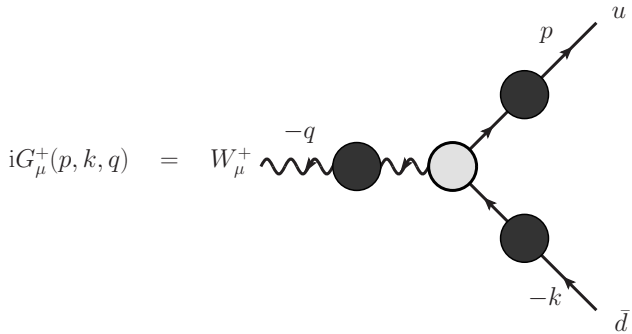}
\caption[]{The diagrammatical representation and momenta assignment of the connected Green's function $\I G_\mu^+(p,k,q)$, Eq.~\eqref{eq:Gamma_general}. The grey blob denotes its 1PI part, $\I \Gamma_\mu^+(p,-k)$, while the dark blobs represent the full propagators. (Notice the arrows on the boson line: We conventionally define the $W^+$ as an antiparticle.)}
\label{fig:vertex}
\end{center}
\end{figure}

Let us work out the idea in detail. Consider the connected Green's function $\langle u^\prime \bar d^\prime W_\mu^+ \rangle$ and define its Fourier transform $\I G_\mu^+(p,k,q)$ as
\begin{eqnarray}
\label{eq:Gamma_general}
\int \d^4 x \, \d^4 y \, \d^4 z \,
\e^{\I p\cdot x}\,\e^{\I k\cdot y}\,\e^{-\I q\cdot z}
\bra{0}T \operator{u^\prime(x)}\operator{\bar d^\prime(y)}\operator{W_\mu^+(z)}\ket{0}
\nonumber
&&
\\
=\, (2\pi)^4\,\delta^4(p+k-q)\,\I G_\mu^+(p,k,q) \,. \quad&&
\end{eqnarray}
For the assignment of the momenta see Fig.~\ref{fig:vertex}. Recall that a connected Green's function is generally calculated as a proper (1PI) Green's function with full propagators at the external lines:
\begin{equation}
\I G_\mu^+(p,k,q)
\>=\>
\operator{\I S_u(p)} \I\Gamma_\nu^+(p,-k) \operator{\I S_d(-k)} \operator{\I D^{\nu}_{\hphantom{\nu}\mu}(q)} \,.
\end{equation}

For the external fermionic lines we consider the full propagators $S_u(p)$, $S_d(p)$, as defined by Eqs.~\eqref{eq:S_u_d}. The $W^\pm$ propagator $D_{\mu\nu}(q)$ is taken at this moment to be just the bare propagator of a massive vector field with hard mass $M_W$. Similarly, the proper vertex $\Gamma_\mu^+(p,-k)$ is taken to be the tree one, determined by the charged current Lagrangian $\eL_{\mathrm{cc}}(u^\prime,d^\prime)$, \eqref{eq:L_gauge_cc}, i.e.,
\begin{equation}
\label{eq:vertex_bare}
\Gamma_\mu^+(p,-k) \>=\> \frac{g}{\sqrt{2}}\gamma_\mu P_L \,.
\end{equation}
(A momentum dependence of gauge boson propagator and the vertex will be discussed in Sec.~\ref{sec:other_momentum_dep}.) Thus, we have at the leading order in the gauge coupling constant $g$ immediately
\begin{equation}
\label{eq:Gamma_explicit}
\I G_\mu^+(p,k,q)
\>=\>
\operator{\I S_u(p)} \I\frac{g}{\sqrt{2}}\gamma^\nu P_L \operator{\I S_d(-k)} \operator{\I D_{\nu\mu}(q)} \,.
\end{equation}


We are now ready to apply the LSZ reduction formula. Recall that upon taking the limit $p^2 \rightarrow m_{u_i}^2$, $k^2 \rightarrow m_{d_j}^2$, $q^2 \rightarrow M_{W}^2$ in the Green's function $\I G_\mu^+(p,k,q)$, the residue of the leading divergent term (i.e., the one with the triple pole) is (up to polarization vectors and a sign) the desired matrix element $\mathcal{M}_{fi}$ of the process $W^+ \rightarrow u_i + \bar d_j$ (cf. Eq.~\eqref{eq:definition:Mfi}). Taking into account the explicit form \eqref{eq:Gamma_explicit} of $\I G_\mu^+(p,k,q)$ and applying the asymptotic formul{\ae} \eqref{eq:S_asymptotics} for the fermionic propagators $S_u(p)$, $S_d(p)$, we arrive straightforwardly at the result
\begin{eqnarray}
\label{eq:LSZ}
&&
\I G_\mu^+(p,k,q) \>
\xrightarrow[\begin{subarray}{c}
p^2 \rightarrow m_{u_i}^2 \\
k^2 \rightarrow m_{d_j}^2 \\
q^2 \rightarrow M_{W}^2 \\
\vphantom{x}
\end{subarray}]{}
\\ &&
\quad\quad\quad
-
\frac{\I\mathcal{U}_{u_i}(p)}{p^2-m_{u_i}^2}
\frac{\I\mathcal{\bar V}_{d_j}(k)}{k^2-m_{d_j}^2}
\frac{\I\varepsilon_\mu^*(q)}{q^2-M_{W}^2}
\,\I\mathcal{M}_{fi}
\>\> + \> \ldots \,,
\nonumber
\end{eqnarray}
where the ellipsis represents less divergent terms (i.e., the terms with double and single poles and regular terms). The matrix element $\mathcal{M}_{fi}$ in \eqref{eq:LSZ} comes out as
\begin{equation}
\label{eq:Mfi_exact}
\mathcal{M}_{fi}
\>=\>
\frac{g}{\sqrt{2}}\operator{\bar u_{u_i}(p)} \big(\tilde V_{u\vphantom{d}}^{\vphantom{\dag}} \tilde V_d^\dag \big)_{ij} \gamma^\mu P_L \operator{v_{d_j}(k)} \operator{\varepsilon_\mu(q)}
\,.
\end{equation}
For the definition of the matrices $\tilde V_u$, $\tilde V_d$ see Sec.~\ref{sec:formalism}, Eq.~\eqref{eq:def:U_V_tilde}.

We are now going to compare the matrix element $\mathcal{M}_{fi}$ with the matrix element $\mathcal{M}_{fi}^{\mathrm{(SM)}}$ (Eq.~\eqref{eq:Mfi_SM}), calculated within the SM for the same process $W^+ \rightarrow u_i + \bar d_j$ and in the same (lowest) order in the gauge coupling constant. Demanding that both matrix elements have the same form, we conclude that the effective CKM matrix is given by
\begin{equation}
\label{eq:CKM_eff}
V_{\mathrm{CKM}}^{\mathrm{(eff)}} = \tilde V_{u\vphantom{d}}^{\vphantom{\dag}} \tilde V_d^\dag \,.
\end{equation}
(We discuss the precise meaning of the adjective \qm{effective} in Sec.~\ref{sec:summary_discussion}.) This effective CKM matrix has the striking feature of being in general non-unitary, in contrast to the CKM matrix \eqref{eq:CKM_SM} in the SM. This is due to non-unitarity of the matrices $\tilde V_u$, $\tilde V_d$. Note, however, that unitarity of $V_{\mathrm{CKM}}^{\mathrm{(eff)}}$ is restored in the special case of constant self-energies $\Sigma_u$, $\Sigma_d$. This is not surprising, since in such a case the exact theory is actually identical to the SM (Sec.~\ref{sec:SM}) and both expressions for the CKM matrix \eqref{eq:CKM_SM} and \eqref{eq:CKM_eff} coincide.

Let us now proceed to the definition of the effective Lagrangian. The CKM matrix in the SM occurs not only in the matrix elements of the type \eqref{eq:Mfi_SM} (in the same way as our effective CKM matrix does), but it also lives in the charged-current Lagrangian \eqref{eq:L_gauge_SM_cc}, written in terms of the mass-diagonalized quark fields $u$, $d$. The natural question arises whether and to what extent it is analogously possible to replace the exact theory by a Lagrangian written in terms of the fields $u$, $d$, which incorporates in a natural way the effective CKM matrix obtained above. The answer is that it is possible merely in an effective sense to be specified below.

Let us define the effective Lagrangian $\eL_{\mathrm{eff}}(u,d)$: We postulate the operators $u$, $d$ in such a way that they are operators of the massive states of the exact theory. More precisely, $\eL_{\mathrm{eff}}(u,d)$ contains, apart from the fermionic kinetic terms, the mass Lagrangian $\eL_{\mathrm{eff,mass}}(u,d)$ of the form
\begin{equation}
\label{eq:L_eff_mass}
\eL_{\mathrm{eff,mass}}(u,d) = - \bar u M_u u - \bar d M_d d \,.
\end{equation}
Here the mass matrices $M_u$, $M_d$ (not to be confused with also diagonal, but momentum-dependent matrices $M_u(p^2)$, $M_d(p^2)$, Eq.~\eqref{def:M_diag_nonconst}) are given by
\begin{subequations}
\begin{eqnarray}
M_u &=& \diag(m_{u_1},m_{u_2},\ldots,m_{u_n}) \,,
\\
M_d &=& \diag(m_{d_1},m_{d_2},\ldots,m_{d_n}) \,,
\end{eqnarray}
\end{subequations}
with the entries determined by the poles of the full propagators $S_u(p)$, $S_d(p)$. Let the effective Lagrangian $\eL_{\mathrm{eff}}(u,d)$ contain also the kinetic terms of the gauge bosons $W^\pm$, $Z$, $\gamma$ and the corresponding mass terms. Since $\eL_{\mathrm{eff}}(u,d)$ is written in terms of massive fields, it has potential to describe processes like $W^+ \rightarrow u_i + \bar d_j$ directly, without employing the LSZ reduction formula. Indeed, postulating that $\eL_{\mathrm{eff}}(u,d)$ contains the SM-like charged-current interactions of the form
\begin{equation}
\label{eq:L_eff_cc}
\eL_{\mathrm{eff,cc}}(u,d) = \frac{g}{\sqrt{2}}\bar u \gamma^\mu P_L \tilde V_{u\vphantom{d}}^{\vphantom{\dag}} \tilde V_d^\dag d W_\mu^\dag + \hc \,,
\end{equation}
it is straightforward to see that this leads to same matrix element \eqref{eq:Mfi_exact} as the exact theory. As expected, comparing this effective charged-current interaction Lagrangian with that of the SM \eqref{eq:L_gauge_SM_cc}, we are again lead to the definition of the effective CKM matrix \eqref{eq:CKM_eff}.

Let us finally comment on why we have considered for defining the effective CKM matrix just the process $W^+ \rightarrow u_i + \bar d_j$ and not some other. First note that we could have as well considered any process related to this one by the crossing symmetry and the result would be the same. Second, a natural question arises why not define the effective CKM matrix by some more complicated process, e.g., the annihilation process $W^+ + W^- \rightarrow q_i + \bar q_j$, $q=u,d$. It is, of course, no problem to calculate the matrix element for this process in the same way we did for the decay process above. However, it turns out to be difficult to parameterize the resulting matrix element by a \emph{constant} effective CKM matrix so that the matrix element had the same form as if it was calculated within the SM. Technically, one would have to solve the equations
\begin{subequations}
\begin{align}
\tilde V_{u\vphantom{d}}^{\vphantom{\dag}} \operator{V_d^\dag(q_1^2)} \big(q_1^2-M_d^2(q_1^2)\big)^{-1} \operator{V_d^{}(q_1^2)} \tilde V_u^\dag
& \\
& \!\!\!\!\!\!\!\!\!\!\!\!\!\!\!\!\!\!\!\!\!\!\!\!\!\!\!\!\!
= \>
V_{\mathrm{CKM}} \big(q_1^2-M_d^2\big)^{-1} V_{\mathrm{CKM}}^\dag
\,, \nonumber
\\
\tilde V_d^{\vphantom{\dag}} \operator{V_{u\vphantom{d}}^\dag(q_2^2)} \big(q_2^2-M_u^2(q_2^2)\big)^{-1} \operator{V_u^{}(q_2^2)} \tilde V_d^\dag
& \\
& \!\!\!\!\!\!\!\!\!\!\!\!\!\!\!\!\!\!\!\!\!\!\!\!\!\!\!\!\!
= \>
V_{\mathrm{CKM}}^\dag \big(q_2^2-M_u^2\big)^{-1} V_{\mathrm{CKM}}
\,. \nonumber
\end{align}
\end{subequations}
Note that in these equations one seeks for a momentum-independent $V_{\mathrm{CKM}}$ (but, again, not necessarily unitary), which solves these equations for all momenta $q_1$, $q_2$. This task turns out to be difficult.

\subsection{Electromagnetic and neutral-current interactions}
\label{subsec:nc_em}

In the same way as we probed in the previous subsection the charged-current sector, it is possible to investigate the flavor mixing also in the electromagnetic and neutral-current sectors. Since the procedure is technically completely analogous, we merely state the result. Considering the decay processes $Z \rightarrow q_i + \bar q_j$ and $\gamma \rightarrow q_i + \bar q_j$, $q=u,d$, we arrive at the corresponding effective interaction Lagrangians (to be part of $\eL_{\mathrm{eff}}(u,d)$)
\begin{subequations}
\label{eq:L_eff_nc_em}
\begin{eqnarray}
\eL_{\mathrm{eff,nc}}(u,d) &=& \frac{g}{2 \cos \theta_W}
\\ && \nonumber
\!\!\!\!\!\!\!\!\!\!\!\!\!\!\!\!\!\!\!\!\!\!\!\!\!\!\!\!\!
\!\!\!\!\!
\sum_{q=u,d} \bar q \gamma^\mu \Big[ (v_q+a_q) \tilde V_q^{} \tilde V_q^\dag P_L + (v_q-a_q) \tilde U_q^{} \tilde U_q^\dag P_R \Big] q Z_\mu \,,
\nonumber
\\
\eL_{\mathrm{eff,em}}(u,d) &=& \sum_{q=u,d} e Q_q
\bar q \gamma^\mu \Big( \tilde V_q^{} \tilde V_q^\dag P_L + \tilde U_q^{} \tilde U_q^\dag P_R \Big) q A_\mu \,.
\nonumber \\ &&
\end{eqnarray}
\end{subequations}
We see that, in contrast to their SM counterparts \eqref{eq:L_gauge_SM_nc}, \eqref{eq:L_gauge_SM_em}, these effective Lagrangians exhibit non-trivial flavor mixing. However, as expected, they reduce to those \eqref{eq:L_gauge_SM_nc}, \eqref{eq:L_gauge_SM_em} of the SM with no flavor mixing in the special case of constant self-energies, since then the matrices $\tilde V_q$, $\tilde U_q$ are unitary.

\section{Other sources of momentum dependence}
\label{sec:other_momentum_dep}


In this section we investigate the effects of momentum dependence of the proper vertices and the gauge boson polarization tensors (both of which are depicted in Fig.~\ref{fig:vertex}). This is technically very similar to what was done in Sec.~\ref{sec:general_selfenergies} with fermionic self-energies, so we will be brief. In the following two subsections we perform the analysis separately first for the gauge boson polarization tensors and then for the gauge boson proper vertices with quarks.

\subsection{Gauge boson polarization tensors}

Momentum dependence of the polarization tensors can induce a mixing among the gauge bosons, much like in the case of quarks. This can of course happen only between the two electrically neutral gauge bosons $\gamma$ and $Z$, therefore we will not concern ourselves with the $W^\pm$ bosons here. The treatment of gauge boson mixing is technically similar to that in the fermion sector, so we are not going into a big detail.

Since $\gamma$ and $Z$ can mix, it is convenient to consider two-component field $V_\mu \equiv (\gamma_\mu,Z_\mu)^T$, because the form factor $\Pi(q^2)$ from its (necessarily transverse) polarization tensor $\Pi_{\mu\nu}(q) = ( q^2 g_{\mu\nu}-q_\mu q_\nu ) \Pi(q^2)$ is in general a non-diagonal $2 \times 2$ matrix. Assuming the matrix $\Pi(q^2)$ to be symmetric, we can diagonalize it via an orthogonal transformation $\Pi(q^2) = \op{O^T(q^2)} \op{\Pi_D(q^2)} \op{O(q^2)}$. Here the matrix $\Pi_D(q^2) = \diag \big(\Pi_{D1}(q^2),\Pi_{D2}(q^2)\big)$ is diagonal, while the matrix $O(q^2)$ is orthogonal and we parameterize it as
\begin{eqnarray}
O(q^2) &=&
\left(\begin{array}{rc}
 \cos \theta(q^2) & \sin \theta(q^2) \\
-\sin \theta(q^2) & \cos \theta(q^2) \\
\end{array}\right) \,.
\end{eqnarray}
The relevant part of the full propagator of $V_\mu$ is proportional to $\big[q^2-q^2 \Pi(q^2)\big]^{-1}$, which is thus diagonalized as  $\big[q^2-q^2 \Pi(q^2)\big]^{-1} = \op{O^T(q^2)} \big[q^2-q^2 \Pi_D(q^2)\big]^{-1} \op{O(q^2)}$. We assume that the first pole equation $q^2-q^2 \Pi_{D1}(q^2)=0$ has the only solution $q^2=0$, which corresponds to the massless photon. For the second pole equation $q^2-q^2 \Pi_{D2}(q^2)=0$ we assume the solution $q^2=M_Z^2$.


Invoking the same method utilizing the LSZ reduction formula as before, the amplitudes of the two desired processes $Z / \gamma \rightarrow q_i + \bar q_j$, $q=u,d$, can be extracted from same momentum-space Green's function $\langle V_\mu q \bar q \rangle$ by taking the momentum carried by $V_\mu$ to $q^2=0$ or $q^2=M_Z^2$, respectively. The resulting effective couplings of quarks to the photon and $Z$ are most compactly written as
\begin{eqnarray}
\label{eq:L_eff_nc_em_gauge}
\eL_{\mathrm{eff,nc+em}}(u,d) &=& \sum_{q=u,d} \bigg\{ \nonumber
\\ && \nonumber \hspace{-2.8cm}
\frac{g}{2 \cos \theta_W}
 \bar q \gamma^\mu \Big[ (v_q+a_q) \tilde V_q^{} \tilde V_q^\dag P_L + (v_q-a_q) \tilde U_q^{} \tilde U_q^\dag P_R \Big] q
\\ && \nonumber \hspace{-0.5cm} \times
\big[\sin\theta(0)\,A_\mu+\cos\theta(M_Z^2)\,Z_\mu\big]
\nonumber \\ && \hspace{-2.8cm}
{}+e Q_q
\bar q \gamma^\mu \Big( \tilde V_q^{} \tilde V_q^\dag P_L + \tilde U_q^{} \tilde U_q^\dag P_R \Big) q
\nonumber \\ && \hspace{-0.5cm} \times
\big[\cos\theta(0)\,A_\mu-\sin\theta(M_Z^2)\,Z_\mu\big]
\bigg\} \,.
\end{eqnarray}
Note that this is a generalization of the previous result \eqref{eq:L_eff_nc_em}, which can be obtained from \eqref{eq:L_eff_nc_em_gauge} as a special case of no gauge boson mixing (i.e., when $\theta(q^2)=0$ for all $q^2$).

\subsection{Proper vertex}

In Subsec.~\ref{subsec:cc} while calculating the amplitude of the process $W^+ \rightarrow u_i + \bar d_j$ we approximated the proper vertex $\Gamma_\mu^+(p,-k)$ as the bare one \eqref{eq:vertex_bare}, which in particular means momentum-independent. Now we redo the calculation with including the most general momentum dependence of the vertex.

The most general form of $\Gamma_\mu^+(p,-k)$ can be parameterized, e.g., as follows:
\begin{equation}
\Gamma_\mu^+(p,-k) \>=\> A_\mu(p,k) + p_\mu B(p,k) - k_\mu C(p,k) \,,
\end{equation}
where
\begin{subequations}
\begin{eqnarray}
A_\mu(p,k) &=& \gamma_\mu A_1 + \slashed{p} \gamma_\mu A_2 - \gamma_\mu A_3 \slashed{k} - \slashed{p} \gamma_\mu A_4 \slashed{k} \,, \qquad
\\
B(p,k) &=& B_1 + \slashed{p} B_2 - B_3 \slashed{k} - \slashed{p} B_4 \slashed{k} \,,
\\
C(p,k) &=& C_1 + \slashed{p} C_2 - C_3 \slashed{k} - \slashed{p} C_4 \slashed{k} \,.
\end{eqnarray}
\end{subequations}
Here the form factors $A_i$, $B_i$, $C_i$, $i=1,2,3,4$ are $n \times n$ matrices in the flavor space and, except from being some linear combinations of $1$ and $\gamma_5$ (or equivalently $P_L$, $P_R$), they do not contain any $\gamma$-matrices. Thus, they can depend only on three independent Lorentz-invariant combinations of the two independent momenta $p$, $k$, which can be chosen conveniently as $p^2$, $k^2$ and $q^2=(p+k)^2$. We do not indicate this dependence explicitly.

For the notational purposes it is convenient to introduce now the $x,y$-dependent matrices $\tilde A^\mu(x,y)$, $\tilde B(x,y)$, $\tilde C(x,y)$, with matrix elements given by
\begin{subequations}
\label{eq:def:ABC_tilde}
\begin{eqnarray}
\tilde A_{ij}^\mu(x,y) &=& \Big[ \bar X_u^{} \gamma^\mu A_1 X_d^{} + x X_u^\dag \gamma^\mu A_2 X_d^{}
\nonumber \\ && \hspace{-6mm} {}
+ \bar X_{u\vphantom{d}}^{\vphantom{\dag}} \gamma^\mu A_3 \bar X_d^\dag y + x X_{u\vphantom{d}}^\dag \gamma^\mu A_4 \bar X_d^\dag y \Big]_{ij} \,,
\\
\tilde B_{ij}(x,y) &=& \Big[ \bar X_u^{} B_1 X_d^{} + x X_u^\dag B_2 X_d^{}
\nonumber \\ && \hspace{2mm} {}
+ \bar X_{u\vphantom{d}}^{\vphantom{\dag}} B_3 \bar X_d^\dag y + x X_{u\vphantom{d}}^\dag B_4 \bar X_d^\dag y \Big]_{ij} \,,
\\
\tilde C_{ij}(x,y) &=& \Big[ \bar X_u^{} C_1 X_d^{} + x X_u^\dag C_2 X_d^{}
\nonumber \\ && \hspace{2mm} {}
+ \bar X_{u\vphantom{d}}^{\vphantom{\dag}} C_3 \bar X_d^\dag y + x X_{u\vphantom{d}}^\dag C_4 \bar X_d^\dag y \Big]_{ij} \,.
\end{eqnarray}
\end{subequations}
These quantities are momentum-independent, since the momenta in form factors $A_i$, $B_i$, $C_i$ are evaluated at $p^2=m_{u_i}^2$, $k^2=m_{d_j}^2$ and $q^2=M_W^2$ and similarly the matrices $X_u$, $X_d$, introduced in Sec.~\ref{sec:formalism}, Eq.~\eqref{eq:def:X_tilde}, are evaluated at $p^2=m_{u_i}^2$, $k^2=m_{d_j}^2$, respectively. The symbols $x$, $y$ are quite general, they can stand for a number, for a matrix or even for a differential operator.

Using the same approach based on the LSZ reduction formula as before, we arrive at the amplitude for the process $W^+ \rightarrow u_i + \bar d_j$, which can be written with the help of notation \eqref{eq:def:ABC_tilde} compactly as
\begin{eqnarray}
\mathcal{M}_{fi}
&=&
\operator{\bar u_{u_i}(p)} \big[ \tilde A^\mu(m_{u_i},m_{d_j}) + p^{\mu} \tilde B(m_{u_i},m_{d_j})
\nonumber \\ && {}
- k^\mu \tilde C(m_{u_i},m_{d_j}) \big]_{ij} \operator{v_{d_j}(k)} \operator{\varepsilon_\mu(q)}
\,.
\end{eqnarray}
This is just a generalization of the result \eqref{eq:Mfi_exact} obtained earlier. Analogously we can generalize the effective charged-current Lagrangian \eqref{eq:L_eff_cc}:
\begin{eqnarray}
\eL_{\mathrm{eff,cc}}(u,d) &=&
\bar u \big[ \tilde A^\mu(-\I \vecleft{\slashed{\partial}},\I \vecright{\slashed{\partial}}) - \I \vecleft{\partial}^{\mu} \tilde B(-\I \vecleft{\slashed{\partial}},\I \vecright{\slashed{\partial}})
\nonumber \\ && {}
+ \I \vecright{\partial}^\mu \tilde C(-\I \vecleft{\slashed{\partial}},\I \vecright{\slashed{\partial}}) \big] d W_\mu^\dag + \hc
\end{eqnarray}
Here $\vecleft{\partial}^{\mu}$ and $\vecright{\partial}^\mu$ operate on $\bar u$ and $d$, respectively

\section{Discussion}
\label{sec:summary_discussion}


Although not expressed by a single massive Lagrangian, we showed in Sec.~\ref{sec:general_selfenergies} that the exact theory allows to calculate amplitudes of the processes involving the massive quark asymptotic states, using the ordinary perturbation theory and the LSZ reduction formula. In this respect the introduction of the effective Lagrangian $\eL_{\mathrm{eff}}(u,d)$ (Eqs.~\eqref{eq:L_eff_mass}, \eqref{eq:L_eff_cc}, \eqref{eq:L_eff_nc_em}) is unnecessary. However, we introduced it since it provides a convenient basis for discussing some of the substantial differences between the general case of the momentum-dependent self-energies and the special, familiar case of the constant self-energies (i.e., the mass matrices).

First, a few comments are in order concerning the effectiveness of $\eL_{\mathrm{eff}}(u,d)$.
It is not effective in the usual sense as being a low energy approximation of some underlying theory, i.e., in our case the exact theory. Rather, it is by construction effective in the sense that it reproduces predictions of the exact theory, but only for a very limited set of processes (and only at the tree level).
Namely, only the processes $W^+ \rightarrow u_i + \bar d_j$ and $Z / \gamma \rightarrow q_i + \bar q_j$, $q=u,d$, modulo crossing symmetry, are computed correctly (i.e., in accordance with the exact theory). If one calculates any more complicated process (e.g., $W^+ + W^- \rightarrow q_i + \bar q_j$) within this effective theory, one obtains an answer differing from the answer obtained within the exact theory. Clearly, we have lost some amount of the physical information contained in the exact theory when passing to the effective one. However, this makes sense, since the self-energies as the momentum-dependent matrix \emph{functions} (the exact theory) contain \qm{much more} physical information than the \emph{constants} like the masses and the flavor mixing matrices (the effective theory).

There is a significant exception, though. In the case of constant self-energies the amount of physical information remains the same while going from the exact theory to the effective one. Recall that in this case the effective Lagrangian $\eL_{\mathrm{eff}}(u,d)$ (Eqs.~\eqref{eq:L_eff_cc}, \eqref{eq:L_eff_nc_em}) reduces precisely to the SM Lagrangian $\eL_{\mathrm{SM}}(u,d)$ (Eq.~\eqref{eq:L_gauge_SM}), which is indeed fully physically equivalent to the original exact theory.

This leads us to another substantial difference between the two cases. We are accustomed from the SM that the interaction basis ($u^\prime$, $d^\prime$) and mass basis ($u$, $d$) are related to each other by a unitary transformation \eqref{eq:def:quark_unitary_redef} and working in either of them is merely a matter of taste. This is clearly not the case in the more general situation of momentum-dependent self-energies: Here the interaction basis operators $u^\prime$, $d^\prime$ are the fundamental ones and there is no way to obtain from them the mass basis operators $u$, $d$ by a suitable unitary transformation. This is of course related to the effective nature of the corresponding Lagrangian $\eL_{\mathrm{eff}}(u,d)$, since the operators $u$, $d$ are nothing more that merely postulated, effective fields.

There is also another way of seeing that the two sets of operators $u^\prime$, $d^\prime$ and $u$, $d$ should be viewed as essentially unrelated. Note that the dimension of the vectors $u$, $d$ is in this paper by construction equal to the number of poles of the corresponding propagators $\langle u^\prime \bar u^\prime \rangle$, $\langle d^\prime \bar d^\prime \rangle$. However, although we did not assume it here for the sake of simplicity, it is in principle thinkable that this number could be different from what we called $n$ -- the dimension of vectors $u^\prime$, $d^\prime$ (cf. the discussion of number of propagator's poles in Sec.~\ref{sec:formalism}). Thus, in such situation it would be obviously impossible to relate the two bases $u^\prime$, $d^\prime$ and $u$, $d$ with \emph{different} dimensions by any regular linear transformation.

The comparison with the Refs.~\cite{Machet:2004rv,Duret:2006wk} is in order now. We have confirmed the phenomenological results concerning non-unitarity of the effective CKM matrix and occurence of the flavor changing electromagnetic and neutral currents. In particular, we recovered the explicit formula \eqref{eq:CKM_eff} for the former. What is new in our treatment is that we provided also explicit formul{\ae} \eqref{eq:L_eff_nc_em} for the flavor mixing in the electromagnetic and neutral-current sectors. Moreover, we found out that the corresponding mixing matrices are only effective ones: They allow us to compute the processes in the lowest order in the gauge coupling constants, but if one wants to go to higher orders of the perturbation theory, it is necessary to come back to the self-energies and consider their full momentum dependence.

We also contributed to the discussion of the relation between the interaction basis $u^\prime$, $d^\prime$ and the mass basis $u$, $d$. We confirmed that both bases cannot be related by a unitary transformation. The authors of Refs.~\cite{Machet:2004rv,Duret:2006wk} showed, however, that the two bases can be related by the \emph{non-unitary} transformation
\begin{equation}
\label{eq:def:quark_unitary_redef_tilde}
u = \tilde X_u^\dag u^\prime \,,
\quad\quad\quad
d = \tilde X_d^\dag d^\prime \,,
\end{equation}
(cf. Eq.~\eqref{eq:def:quark_unitary_redef}) with non-unitary $\tilde X$'s defined by Eq.~\eqref{eq:def:X_tilde}. (The resulting non-diagonality of quark kinetic terms due to non-unitarity of matrices $\tilde X$ can be cured by adding appropriate finite counterterms to Lagrangian \cite{Shabalin:1978rs,Duret:2008md,Duret:2008st}.) This is in accordance with our result: Using the non-unitary redefinitions \eqref{eq:def:quark_unitary_redef_tilde} in the Lagrangian \eqref{eq:L_gauge_SM} (and neglecting impacts on kinetic terms), we arrive precisely at our effective Lagrangian \eqref{eq:L_eff_cc}, \eqref{eq:L_eff_nc_em}. Since we argued, however, that any Lagrangian, written in the mass basis, should be (at least in principle) regarded as an effective one, in the sense described above, we conclude that the non-unitary transformations \eqref{eq:def:quark_unitary_redef_tilde} should be regarded as effective as well.

\section{Summary and conclusion}
\label{sec:conclusion}

The EWSB mechanism employed in the SM -- the Higgs mechanism -- manifests itself by generation of the fermionic mass matrices in the Lagrangian. We considered in this paper a more general case: Instead of constant mass matrices we assumed that some unknown EWSB mechanism generates full symmetry-breaking fermionic propagators, with 1PI parts (the proper self-energies) given by flavor matrices with generic momentum dependence. Restricting ourselves only to the quark sector, we investigated some consequences of this generalization.

We computed amplitudes for some elementary processes involving the massive quarks and arrived at substantially different predictions from those of the SM. In the charged-current sector, we managed to parameterize the relevant amplitudes by an effective CKM matrix, which was a general \emph{non-unitary} matrix. Recall that the unitarity of the CKM matrix in the SM is established theoretically \cite{Cabibbo:1963yz,Kobayashi:1973fv} as well as experimentally \cite{Tarantino:2009sx,Amsler:2008zzb}. In the electromagnetic and charged-current sectors we found that the relevant processes change flavor at the leading order in the gauge coupling constants. Again, it is a well established theoretical \cite{Glashow:1970gm} and experimental \cite{Amsler:2008zzb} fact that the flavor changing electromagnetic and neutral-current interactions are strongly suppressed.

All of these phenomenological flaws are obviously consequences of non-trivial momentum dependence of the quark self-energies, because in the special case of constant self-energies we just obtain the SM. Since the particular shape of the self-energies' momentum dependence depends on a particular model of EWSB, we have in this manner provided a way leading to a possibility of discriminating among various alternative models of EWSB.

Referring to the properties of the CKM matrix in the SM, it is worth spending here a few words concerning its perturbative renormalization. We have considered throughout the paper only the constant self-energies (mass matrices) in the SM, in which case the unitarity of the CKM matrix is obviously guaranteed. However, there is a natural question whether this proclaimed unitarity remains preserved upon the radiative corrections and corresponding infinite renormalization. This question was discusses in the literature by various authors (e.g., \cite{Barroso:2000is,Diener:2001qt,Zhou:2003te,Kniehl:2006rc}), using a variety of (on-shell) renormalization schemes for the CKM matrix. Carrying on the renormalization within a general $R_\xi$ gauge, they showed that as long as one requires that the renormalized CKM matrix be gauge independent, its unitary stays indeed preserved. Moreover, they showed that the radiative corrections to the CKM matrix are negligible, of order $10^{-5}$ \cite{Denner:1990yz,Kniehl:2000rb}, and are therefore of little phenomenological importance. In fact, all experimental determinations \cite{Amsler:2008zzb} of its elements are based on the (bare) CKM matrix of the form considered in Sec.~\ref{sec:SM}.

Apart from the phenomenological consequences, we showed that the self-energies' momentum dependence has also some conceptual impacts on the way we are accustomed to understand some basic notions in the SM. In particular, we argued that the interaction basis and the mass basis should be in principle regarded as essentially unrelated (e.g., by a unitary transformation), since the latter one is only an appropriately postulated effective notion.
Noting that such conceptual subtleties are not an issue in the SM, on should still keep in mind that this is (at least from the point of view of the present analysis) merely an \qm{accidental} consequence of the employed particular mechanism of EWSB in the SM.

\begin{acknowledgments}
The author is grateful to Ji\v{r}\'{\i} Ho\v{s}ek, Tom\'{a}\v{s} Brauner, Adam Smetana, and Hynek B\'{\i}la for reading the manuscript and for useful comments and to Bruno Machet for enlightening discussions.
\end{acknowledgments}


\end{document}